\documentclass[11pt,twoside,a4paper]{article}

\usepackage[american]{babel}
\usepackage{bm}
\usepackage{amsfonts}
\usepackage{graphicx}
\usepackage{amsmath}
\usepackage{fancyhdr}

\usepackage{setspace}
\newcommand{\nn}{\nonumber}

\usepackage{color}

\newcommand{\zd}{\delta}

\newcommand{\zs}{\sigma}

\newcommand{\ze}{\varepsilon}

\newcommand{\zg}{\gamma}
\newcommand{\zl}{\lambda}

\newcommand{\zo}{\omega}
\newcommand{\zr}{\rho}

\newcommand{\dif}{\; \textrm d }

\newcommand{\dpp}[2]{\frac{  \partial #1 }{  \partial #2   } }

\begin{document}

\begin{center}
\Large{\textbf{Study of the ultrafast dynamics of ferromagnetic materials with a Quantum Monte Carlo atomistic model}}\\
\vskip0.5cm
	\small{O. Morandi, P.-A. Hervieux}	\\
\vskip0.5cm
	\textit{\textsf{Institut de Physique et Chimie des Mat\'eriaux de Strasbourg \\
23, rue du Loess,
F-67034 Strasbourg, France}}	
\vskip0.5cm
	\textit{omar.morandi@unifi.it}
\end{center}
\begin{center}
\begin{minipage}[h]{0.8\textwidth}
\section*{\textsf{Abstract}}
 \small
\textsf{
We study of the ultrafast dynamics of the atomic angular momentum in ferrimagnets irradiated by laser pulses. My apply a quantum atomistic spin approach based on the Monte Carlo technique. Our model describes the coherent transfer of angular momentum between the spin and the orbital momentum as well as the quenching of the orbital momentum induced by the lattice field. The Elliott-Yafet collision mechanism is also included. We focus on elementary mechanisms that lead to the dissipation of the total angular momentum in a rare earth-transition metal (RE-TM) alloy in which the two sublattices have opposite spin orientation. 
Our model shows that the observed ultrafast quenching of the magnetization can be explained  microscopically by the transfer of spin between the sublattices and by the quenching of the  localized orbital angular momentum.  
}
\end{minipage}
\end{center}
\vspace{0.5cm}
\normalsize

\section{Introduction}
\label{sec:intro}

The study of the out-of-equilibrium properties of strongly excited magnetic systems is today a very active area of research. Understanding the physics of the ultrafast magnetization dynamics in clusters of interacting molecules or solids constitutes an important challenge both from the experimental and theoretical point of view.
Since the discovery by Beaurepaire et al. \cite{Beaurepaire_96} in 1996 of the femtosecond ultrafast demagnetization in magnetic metals induced by laser, the elementary mechanisms which are at the basis of such a magneto-optical phenomena are at the centre of an intense debate.

In the framework of the phenomenological three temperature model, the atomic spin, the kinetic energy of electrons and the phonon bath are considered as three interacting reservoirs of energy and angular momentum. The infrared femtosecond laser pump pulse injects energy in the magnetic material by perturbing the electronic distribution in the vicinity of the Fermi level \cite{Hohlfeld_01,Bonn_00}. The excited electrons transfer energy to the lattice via collisions with phonons and to the spins in a time scale of hundred of femtoseconds leading to the so-called ultrafast demagnetization. Nevertheless, this model does not address the specific question of angular momentum conservation at the femtosecond time scale. Despite intense theoretical and experimental efforts, no consensus has been found in the community concerning the microscopic mechanisms that allows transferring angular momentum away from the spin degree of freedom \cite{Kyriluyk_10}.

Several mechanisms have been proposed such as coherent interaction between the electron spins and electrical field of the laser \cite{Zhang_00,Bigot_09}, electrons-magnons scattering \cite{Carpene_09}, defect and phonons induced Elliot-Yafet spin flips \cite{Koopmans_05,Koopmans_10}, ultrasfast quenching of the magneto crystalline anisotropy \cite{Boeglin_10} and superdiffusive spin transport \cite{Battiato_10,Battiato_12}. Even if all those mechanisms were sustained by experimental evidences \cite{Melnikov_11,Sultan_12,Wieczorek_15}, interrogations remain concerning their actual efficiency \cite{Carva_11,Schellekens_13}.
Recently, the ultrafast demagnetization in some bulk transition metal has been modeled by the time dependent Kohn-Sham  density functional theory \cite{Krieer_15}. It is found that the spin-orbit interaction plays a central role in the description of the demagnetization dynamics. However, due to the high computational cost required by such calculations only simple structures like for example the bulk materials can be simulated. Moreover, the inclusion of the phonon scattering processes goes beyond the up-to-date calculation possibilities.


Anti-ferromagnetic alloys are prominent materials for the theoretical and experimental study of the transfer of spin and angular momentum between valence and conduction electrons. They are characterized by  two or more sublattices with opposite spin polarization. The most investigated anti-ferromagnetic alloys are realized by bounding together transition metals (TM) (typically Fe \cite{Radu_11,Wienholdt_13}, Co \cite{Lopez_12} and Pd \cite{Vodungbo_12}) and  rare earth (RE) atoms (Gd \cite{Graves_13,Wienholdt_13,Radu_11} and Tb \cite{Bergeard_14}).
Since the transfer of spin between the two sublattices activated by the laser field turns out to be a particularly efficient process, such materials have very short demagnetization times.

It is well known that the magnetization quenching in pure TM  \cite{Beaurepaire_96,Stamm_07} and pure RE  \cite{Wietstruk_11} films excited by fs laser pulses occurs on different characteristic times \cite{Radu_15}. This difference has been attributed to the nature of the orbitals that carry the magnetic moment: the itinerants 3d electrons of TM are directly excited by the laser pulse while in the case of 4f localized electrons the excitation is mediated though the 5d-4f exchange coupling \cite{Andres_15}. 
Element resolved investigation of ultrafast magnetization dynamics in TM-RE alloys gives a unique opportunity to study in details the transfer of angular momentum in coupled systems. So far, atomistic models have been proposed in order to reproduce the ultrafast magnetization dynamics in RE-TM \cite{Radu_11,Schellekens_13,Wienholdt_13,Mentink_14}. However, the numerical results are strongly affected by the values of the exchange constants that are usually free parameters. Moreover, the itinerant character of the d-band electrons cannot be reproduced by such an approach.
In fact, atomistic models are based on the classical description of the atomic spin and angular momentum provided by the phenomenological Landau-Lifshitz-Gilbert (LLG) equation. The use of the LLG equations for the description of the electron spin raises some questions concerning the validity a classical approach to the study of the magnetic properties of nanometric systems.


In the absence of dissipations the LLG equations conserve the total angular momentum of the system. The sources of
energy dissipation such as the couplings with the phonons and with the free electrons are modeled by a fictitious external
magnetic field which has the characteristic of a thermal noise. This introduces in the model a uncontrolled source of
loss of the total angular momentum. 


In this Contribution, we propose a new atomistic-like approach where a system constituted of interacting atoms is described in a quantum framework. Our model  describes the coherent exchange of spin and orbital momentum among the different species of atoms as well as the sources of dissipation of the angular momentum (interaction with the phonon bath and orbital quenching). By identifying the channels of dissipation of the atomic angular momentum and the efficiency of the spin exchange process, our model shares light on the microscopic dynamics which is at the origin of the evolution of the magnetization observed recently in some nanomaterials \cite{Bergeard_14}.

\section{Model}

The magnetic configuration of a complex material like the RE-TM alloys results from the interaction between localized and itinerant electrons. The magnetism of the RE atoms is essentially due to the
localized f orbitals, while the magnetic properties of the TM sublattice arise from the d orbitals that have a mixed
localized-itinerant character.

Our model describes localized orbitals as well itinerant electrons.
In order to illustrate our method, we consider a prototype of the RE-TM material consisting only by two  types of orbitals,
 the localized (f-type) and the itinerant (d-type) orbitals.
The generalization of the model to  more realistic situations described in the final part of the paper is straightforward.

It is convenient to distinguish between interatomic and intra-atomic processes.
We denote by interatomic processes all the phenomena in which localized and delocalized electrons exchange spin
 and energy with the surrounding atoms and with the phonon bath. The remaining (intra-atomic) processes describe
the local exchange of spin and angular momentum due to the spin orbit interaction and the orbital quenching.

Firstly, we focus on the intra-atomic processes. Various experiments and theoretical models indicate that
the spin-orbit interaction
may play a central role on the evolution of the atomic spin \cite{Bigot_13,Krieer_15,Mueller_13,Carva_11}.
In fact, microscopic mechanisms such as the Elliott-Yafet, Dyakonov-Perel and the Rashba effect, which are
responsible of the dissipation of the total angular momentum of the electrons, are only
different manifestation of the spin-orbit interaction.
We focus here on the theoretical modeling of the out-of-equilibrium transfer of momentum between spin and orbital degree of freedom.
This issue is rather unexplored.

The intra-atomic Hamiltonian 
takes the form
\begin{align}
\mathcal{H}(\mathbf{R}_i)= \zl_{SO} \mathbf{L}_i \cdot \mathbf{S}_i+  \mathbf{S}_i \cdot \sum_{\langle j\in \textrm{ NA}_i \rangle} \zg_{ij}   \mathbf{S}_j  + V_l\;. \label{atom_ham}
\end{align}
Here, $\mathbf{R}_i$ denotes the lattice position. The first term of the equation is the spin-orbit interaction. It is responsible of the mixing of spin and orbital momentum.
The coefficient $ \zl_{SO}$ is the spin-orbit strength and $\mathbf{S}_i$,  $\mathbf{L}_i $ are, respectively,
the local spin and orbital momentum operators.  The second term of Eq. \eqref{atom_ham} is the spin exchange interaction. This term indicates that the atom at the position $\mathbf{R}_i$ feels an effective magnetic field proportional to the values of
the  spin of the atom at $\mathbf{R}_j$ weighed by the exchange interaction coefficient $\gamma_{ij}$. The  sum runs over the neighbors  atoms (NA).

The last term of Eq. \eqref{atom_ham} describes the electrostatic  crystal field. It is responsible of the quenching of the atomic angular momentum. The spin-orbit interaction tends to align the spin and the orbital motion of the electrons. 
This mechanism competes with the orbital quenching due to the crystal field. In fact, the spherical symmetry of the atomic potential is broken by the lattice potential. The angular momentum is no longer a good quantum number and the atomic wavefunction have a mixed character. 
In many 3d ions the crystal field interaction is much stronger than the spin-orbit interaction and the orbital momentum is completely quenched (orbital momentum $L_z=0$).
However, in higher transition metal ions (the 4d and 5d series) the effects of the crystal field and the spin-orbit interaction can be comparable and the orbital momentum may not be completely quenched.

We model the crystal lattice potential by taking the first correction to the spherical potential for a cubic lattice $V_l= \zl_{l}(x^4+y^4+z^4)$. The strength of the molecular quenching field is given by the parameter $\zl_{l}$. 

The only interaction of the $i$-th atom with the surrounding atoms in Eq. \eqref{atom_ham}
arises from the molecular exchange field $\sum \zg_{ij}    \mathbf{S}_j  $.
In order to reproduce the full many-body dynamics, we complete the description of the system by including
the processes of dissipation of energy and angular momentum of the atoms
with the itinerant electrons and the phonons (interatomic processes).

We represent the atom wave function in the product space of spin and orbital momentum evaluated along a fixed quantization axis. We denote by $\zr_{m,l;m',l'}(\mathbf{R}_i)$
the density matrix of the electrons in the localized orbital at the position $\mathbf{R}_i$, where $m$, $m'$ ($l$, $l'$)
denote the spin (orbital momentum).
The relevant equation for $\zr$ is
\begin{align}
 \dpp{\zr_{m,l;m',l'}(\mathbf{R}_i,t)}{t}  =  & -i\hbar \left[\mathcal{H}(\mathbf{R}_i),\zr  \right]+
  \left. \dpp{\zr}{t} \right|_{col} \;. \label{evol_atom_S_L}
\end{align}
The interatomic interactions are described by the last term of the Von-Neumann equation \eqref{evol_atom_S_L}.
We model the interatomic processes in terms of Markov instantaneous collisions.
They are described by two additional Boltzmann master equations for the itinerant and localized electron densities
\begin{align}
\left. \dpp{\zr}{t} \right|_{col} = 
&    \Gamma^\textrm{loc-it}_{m}      -  \frac{n_m  }{\tau^\textrm{loc-it}_{m}}\label{Boltz n}\\
\dpp{f_\zs(\mathbf{k},\mathbf{r})}{t}= &  \Gamma^\textrm{it-loc}_{\zs}     +\Gamma^{\textrm{it-ph}}_{\zs}    -  f_\zs\left(\frac{ 1 }{\tau^\textrm{it-loc}_{\zs}}+\frac{1 }{\tau^{\textrm{it-ph}}_{\zs}}\right) \label{Boltz f}\; .
\end{align}
We denoted by $f_\zs(\mathbf{k},\mathbf{r})$ the density of itinerant electrons at  position $\mathbf{r}$
 with spin $\zs$ ($\zs=\uparrow,\downarrow$) and momentum $\mathbf{k}$,
and by $n_m (\mathbf{R})\equiv \sum_l \zr_{m,l;m,l}$  the $m-$th diagonal element of the atomic density matrix.
In view to the application of a Monte Carlo (MC) solver technique, we have separated the master equations in two parts.
The operators denoted by the symbol $\Gamma$ (gain terms) describe the microscopic processes that increase the local spin density.
The remaining parts, the loss terms, are expressed in terms of a relaxation time $\tau$.
In this way we put in evidence that each interaction processes is associated to a certain collision frequency  $1/\tau$.
The collision frequencies play a central role in the numerical MC scheme applied to Eqs. \eqref{Boltz n}-\eqref{Boltz f}.
The exchange of spin between localized and itinerant electrons is described by the operators $\Gamma^\textrm{loc-it}$ and $\Gamma^\textrm{it-loc}$ for
the localized orbital and the itinerant charges respectively.
\begin{align*}
     \Gamma^\textrm{loc-it}_{m} (\mathbf{R}) =  &   \int_{\textrm{FBZ}}\left( n_{m+1} w_{+}^m   \mathcal{I}^{m,m+1}_{\uparrow,\downarrow}  +   n_{m-1} w_{-}^m \mathcal{I}^{m-1,m}_{\downarrow,\uparrow}  \right)  \frac{\dif\mathbf{k}}{(2\pi)^3}\; ,\\
   %
  \Gamma^\textrm{it-loc}_{\uparrow}  (\mathbf{k},\mathbf{r})  =  &\sum_{m} n_{m+1} \;  w_{-}^m   \mathcal{I}^{m,m+1}_{\uparrow,\downarrow}\; ,\\
  \Gamma^\textrm{it-loc}_{\downarrow}  (\mathbf{k},\mathbf{r})  =  &\sum_{m} n_{m-1} \;  w_{+}^m   \mathcal{I}^{m,m-1}_{\downarrow,\uparrow}\; ,
\end{align*}
where $ w_{\pm}^m  = \frac{\zg^2}{\hbar} \left( S(S+1)- m(m \pm 1)\right)$, $S$ is the total spin, $\zg$ is the exchange interaction between itinerant and bound electrons, FBZ indicates the First Brillouin Zone, and
\begin{align*}
   \mathcal{I}^{m,m'}_{\zs,\zs'} =  &  2\pi    \int_{\textrm{FBZ}}
   [1-f_{\zs }(\mathbf{k})]  f_{\zs'}(\mathbf{k}')       \zd( E_{\zs } (\mathbf{k})-E_{\zs'} (\mathbf{k}')+\ze_m -\ze_{m'}  )\frac{\dif\mathbf{k}'}{(2\pi)^3}\; .
\end{align*}
 We denoted by $E_{\zs } (\mathbf{k}) $ and $\ze_m$ 
 respectively the energy of the itinerant
electrons and   of the localized electrons.
The characteristic times associated to the $\Gamma^\textrm{loc-it}$ and $\Gamma^\textrm{it-loc}$ processes are given by
\begin{align}
  \frac{1}{ \tau^\textrm{loc-it}_{m}}=  &  \underbrace{  w_{-}^m   \int_{\textrm{FBZ}}\mathcal{I}_{m,m+2\zs,\zs,-\zs}
    \frac{ \dif\mathbf{k}}{(2\pi)^3} }_{1/\tau_m^-} \nn \\&+ \underbrace{ w_{+}^m
    \int_{\textrm{FBZ}}\mathcal{I}_{m+2\zs,m,-\zs,\zs}  \frac{  \dif\mathbf{k}}{(2\pi)^3}}_{1/\tau_m^+} \label{tau_f} \\
  \frac{1}{ \tau^\textrm{it-loc}_{\zs}(\mathbf{k})} =  &  \frac{ 1 }{(2\pi)^3}\sum_{m}    n_{m}  \mathcal{I}_{m,m+2\zs,\zs,-\zs}    w_{ \zs }^m  \label{tau_l} \;.
\end{align}
Finally, we include in our model the Elliott-Yafet (EY) electron-phonon  interaction with and without spin conservation  \cite{Shen_14,Baral_14}.
Such collision terms are described by the scattering kernel $\Gamma^{\textrm{it-ph}}$ and $\tau^{\textrm{it-ph}}$. Their expressions are given In Appendix \ref{Sec_app1}.

We solve the coupled von-Neumann Boltzmann  \eqref{evol_atom_S_L}-\eqref{Boltz n} system
by applying a MC approach. We illustrate our MC technique applied to the calculation of the dynamics of the
density matrix $\zr$ colliding with the itinerant electrons.

According to the MC procedure, from Eq. \eqref{Boltz n} we have that the probability that the $i$-th atom
will not experience any collision during the time interval $t$ is  $P=e^{-t/\tau}$ where $\tau^{-1}=\sum_m  1/\tau_m^{f}(\mathbf{R}_i)$ \cite{Jacoboni_book}. By sampling the distribution $P$, we generate a set of the random numbers $t^*$. We interpret $t^*$ as the time at which
the atom at position $\mathbf{R}_i$ collides with an itinerant electron.
We solve the coherent evolution equation for the
$i-$atom (Eq. \eqref{evol_atom_S_L} without the last term) from the initial time $t_0$ to $t^*$.
The diagonal elements of the density matrix
$\zr_{m,l;m,l} (\mathbf{R}_i,t^*)$ give the probability that at the time $t^*$ the $i-$th atomic spin and
angular momentum are  $m$ and $l$ respectively. We model the collision event in $t=t^*$ as the measurement
of the quantum mechanical state of the localized electrons. We select one of all the possible values of
the spin-angular momentum pairs by generating random numbers with probability $\zr_{m,l;m,l} (t^*)$.
We denote by ($m^*,l^*$) the selected values. After the collision, the outcoming state will be
$(m^*+1,l^* )$ or $(m^*-1,l^*)$ with probabilities proportional respectively to $1/\tau_{m^*}^+$ and $1/\tau_{m^*}^-$ of Eq. \eqref{tau_f}.
This procedure is repeated for all the atoms of the system.

\section{Results}

Time-resolved X-ray magnetic circular dichroism techniques allow to resolve the femtosecond spin and orbital angular momentum dynamics of a magnetic nanostructure induced by laser pulses. In the case of composite materials, it is also possible to distinguish the magnetic signal of each sublattices. We simulate the ultrafast evolution of spin and orbital angular momentum in a Co$_{26}$Tb$_{74}$ alloy excited by a femtosecond X-ray laser pulse. We compare our results with the measurements that have been recently performed by Bergeard et al. \cite{Bergeard_14} on the same material.
We simulate a cube of Co$_{26}$Tb$_{74}$  containing around $6\times 10^4$ atoms of Co  and $2\times 10^4$ atoms of Tb.
All the parameters required by our model could be obtained by performing static DFT calculations. The Co$_{26}$Tb$_{74}$ alloy is an amorphous material, the samples show assembly of disordered microcrystals. Under such conditions the DFT calculations become extremely complex and less reliable.
%
%
Methods based on the Quantum MC approach reproduce with good accuracy the static magnetization in solids \cite{Bouzerar_13,Pajda_13}. We found the value of the exchange parameters by fitting the measured static magnetization (Curie temperature 700 K and compensation temperature 500 K \cite{Lopez_13}) of the alloy.  We obtain $\zg_{Co-Tb}=-6$ meV/bound $\zg_{Co-Co}=12$ meV/bound and  $\zg_{Tb-Tb}=1$ meV/bound. 
Concerning the band structure of the delocalized d obitals, we used the profile of the density of states obtained by DFT calculations for pure Co and Tb. Finally, the $\zl$ parameters that appear in Eq. \eqref{atom_ham} are obtained by reproducing the measured static  mean value  of the orbital angular momentum of Co and Tb. We found  $\zl_{l}=25$ meV for   Co and for  Tb $\zl_{l}=1$ meV.  For the spin-orbit interaction we use $\zl^{Co}_{SO}=20$ meV $\zl^{Tb}_{SO}=100$ meV. Such values are in agreement with the values obtained by DFT calculations for perfect  materials.

\begin{figure}[t]
\begin{center}
\includegraphics[width=0.49\columnwidth]{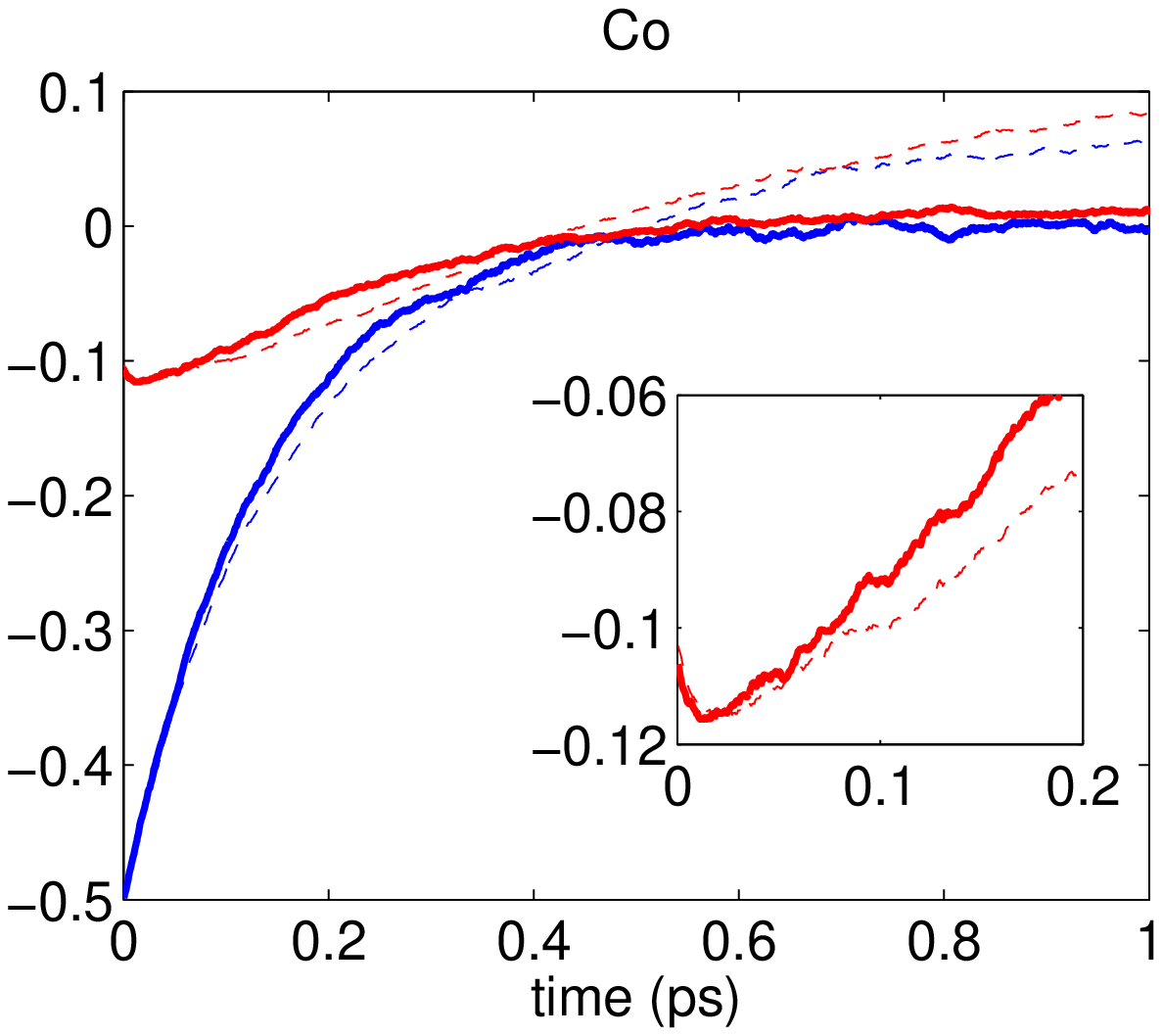}
\includegraphics[width=0.49\columnwidth]{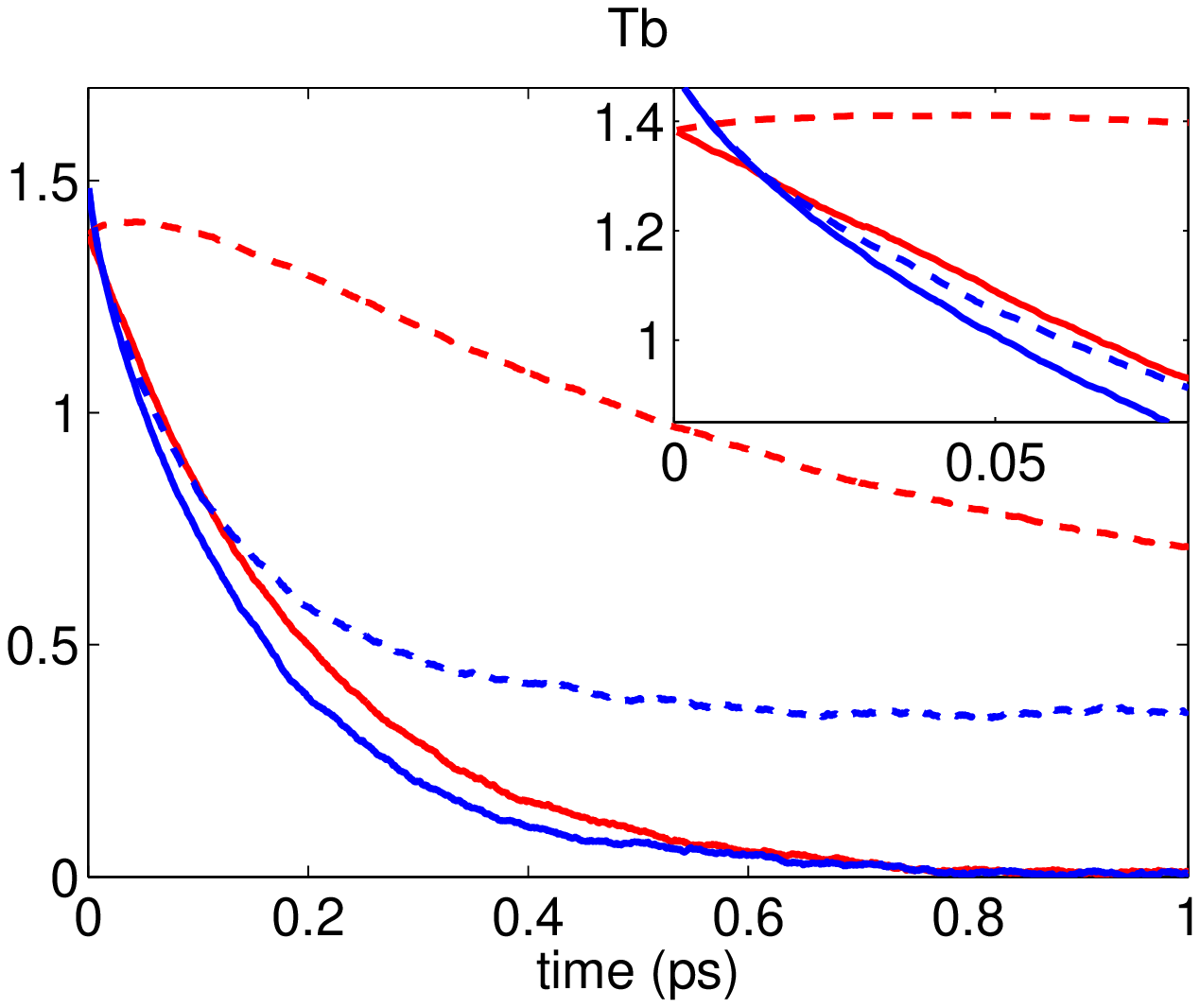}
\caption{Ultrafast evolution of the spin (blue curve) and orbital angular momentum (red curve) after laser excitation for the Cobalt (left panel) and the Terbium (right panel).  The dashed curves are obtained by imposing the conservation of the total angular momentum of the system. The inset in the right panel depicts the zoom of the results for small time.    \label{fig_CoTb_dy2} }
\end{center}
\end{figure}
\begin{figure}[t]
\begin{center}
\includegraphics[width=0.49\columnwidth]{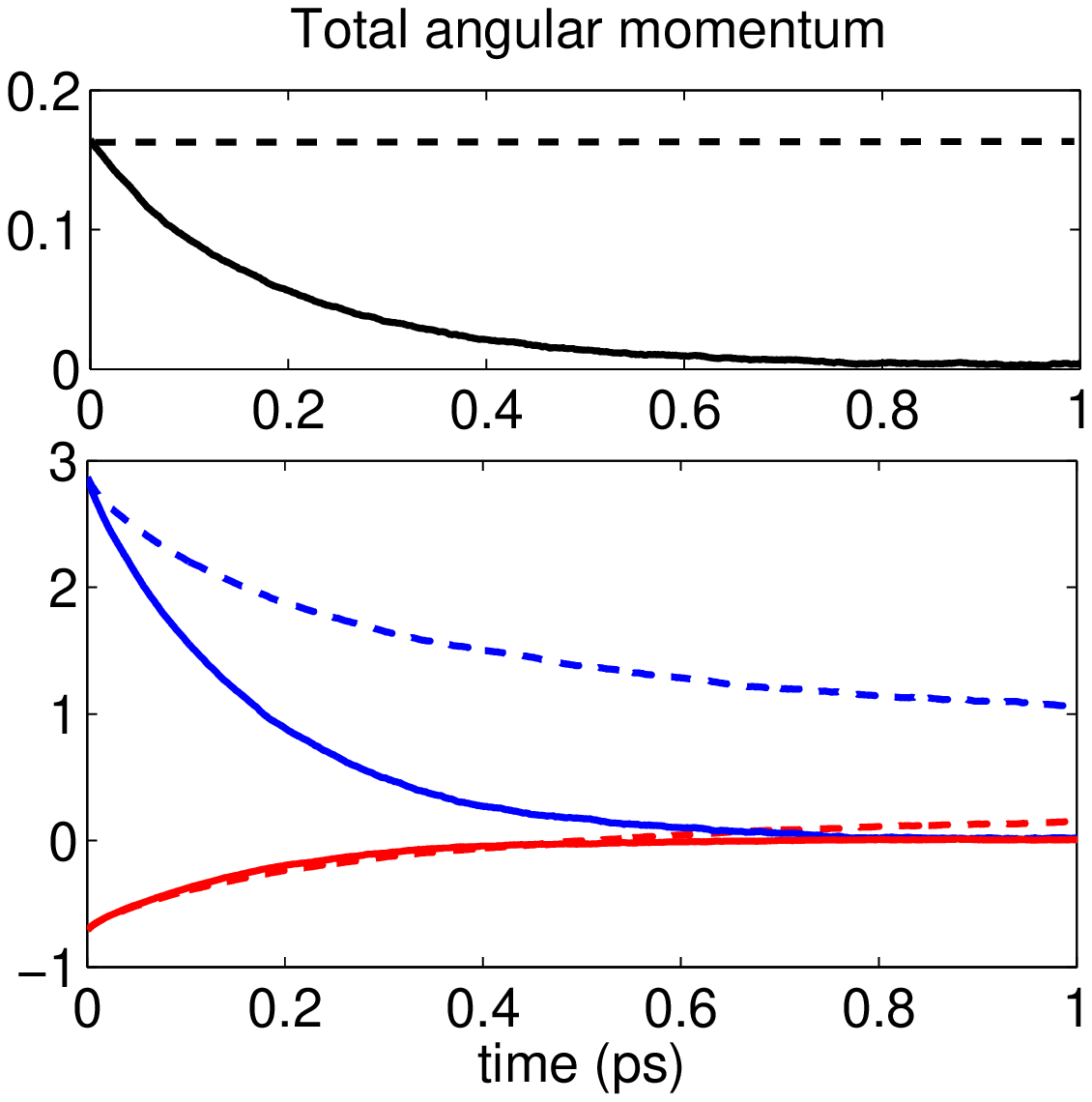}\\
\caption{Upper panel: Total angular momentum of the alloy. Down panel:  Ultrafast evolution of the total angular momentum (spin plus orbital momentum) for Cobalt (red curve) and Terbium (blue curve). The continuous curves refer to the full solution and the dashed curves refer to the case with conservation of the total angular momentum.    \label{fig_CoTb_dy3} }
\end{center}
\end{figure}

We pass now to describe the ultrafast quenching of the magnetization of the alloy induced by a laser pulse. We set the initial configuration of the solid by assuming that for $t=0$ the system is at equilibrium at  temperature 100 K.
Experimental evidences indicate that the energy of the laser field is transferred to the solid by thermal excitation of the TM itinerant electrons. In our approach we discard the direct coherent coupling between the charges and the laser photons \cite{Bigot_09}. We assume that the laser wave raises instantaneously the temperature of the d band electrons of the Cobalt at the temperature of 1500 K that is far above the Curie temperature \cite{Beaurepaire_96,Wienholdt_13}.

The evolution of the spin and orbital momentum of the two sublattices is displayed in Fig. \ref{fig_CoTb_dy2}. The full curves depict the evolution of the total spin (blue curves) and of the orbital momentum (red curves). The left (right) panel refers to the Cobalt (Terbium) sublattice. The evolution of the total angular momentum is depicted in Fig. \eqref{fig_CoTb_dy3}. In the upper panel we display the total angular momentum (spin plus orbital momentum) of the sample and in the bottom panel we discriminate between Cobalt (red curve) and Terbium (blue curve).

In agreement with the experimental results \cite{Bergeard_14}, our simulations show that the laser excitation induces ultrafast transfer of spin between the RE and the TM sublattices. We observe the  quenching of the total magnetization in a time of around 1 ps. Since Terbium and Cobalt are  coupled anti-ferromagnetically, the exchange of spin between the two sublattices leads to a decrease of the magnetism in Cobalt and Terbium that go faster than the quenching of the total magnetization induced by dissipations.

In our simulation, the loss of total angular momentum arises from two physical mechanisms: the EY electron-phonon interaction where the spin of the itinerant electrons is not conserved and the orbital quenching of the localized orbtials modeled by the term $V_l$ in Eq. \eqref{atom_ham}.

By comparing the magnetization dynamics of Tb and Co with the total angular momentum of the system, Bergeard et al. suggested that the RE-TM spin exchange is not necessarily associated to the quenching of the total angular momentum. In order to investigate this statement, we have performed a set of simulations in which we eliminate the  processes that do not conserve the spin. In this case the total angular momentum of the system is exactly conserved.

The results are displayed in Figs. \ref{fig_CoTb_dy2}-\ref{fig_CoTb_dy3} with dashed curves. Our simulations show that during the first 200 fs the spin dynamics of Cobalt and Terbium is essentially the same as the complete simulation. However, the dynamics of the Td orbital angular momentum of the conservative case differs significantly from the results of the full simulations.

The results of our simulations can be interpreted as follows. The laser increases the kinetic energy of the d electrons of  Cobalt. The hot electrons transfer energy to the RE electrons by exchanging spins. This leads to the decreasing of the spin polarization of the localized f electrons of Terbium. The excess of energy passes from the spin to the orbital momentum via the SO interaction. In the real case, where the dissipation of the orbital momentum and spin is included, the orbital momentum is efficiently dissipated by the orbital quenching. Our simulations show that in the case of a conservative system, the orbital momentum initially increases (see insets of Fig. \ref{fig_CoTb_dy2}). This is the signature of the transfer of momentum  between spin and the orbital momentum that behaves as a reservoir of angular momentum.  Such transitory polarization is transferred back to the spin system in a longer time scale.

Experiments show that after laser excitation both spin and orbital angular momentum decreases  \cite{Boeglin_10,Stamm_07}. Consequently, it is typically concluded that despite the presence of the spin-orbit coupling there is not transfer of angular momentum  between spin and angular momentum.
Our simulations help to clarify this point. We showed that the spin-orbit coupling activates the transfer of angular momentum between the spin and orbital degree of freedom in the very early stage of the dynamics. However, due to the quenching of the orbital momentum by the lattice field, the increasing of the orbital momentum cannot be observed within the experimental time resolution.

We have developed a quantum model that describes the time evolution of the spin and orbital angular momentum of d and f electrons in a magnetic composite material. By using a MC approach we describe the coherent atomic evolution due to the spin-orbit, the spin exchange and the orbital quenching as well as the electron-phonon scattering. Our results are in good agreement with the ultrafast dynamics observed in ferrimagnetic alloy. Our model is able to discriminate between the spin and the orbital components of the total angular momentum. We identify the quenching of the local orbital momentum as the main channel of loss of angular momentum during the early stage of the magnetization dynamics.



%

\appendix
\section{Supplementary material: electron-phonon interaction}\label{Sec_app1}
The electron-phonon collision is described by $\Gamma^{\textrm{it-ph}}$ and $\Gamma^{\textrm{it-ph}}$ of Eqs. \eqref{Boltz f}-\eqref{Boltz n}
We write
$
\Gamma^{\textrm{it-ph}}_{\zs}
=      \Gamma^{\textrm{em}}_{\zs}   +\Gamma^{\textrm{ab}}_{\zs}
$
where the acronym  ``em"  and  ``ab" denote respectively the phonon emission and absorption processes. We have (for a similar model Ref. \cite{Schellekens_13})
\begin{align*}
\Gamma^{\textrm{em}}_{\zs}  (\mathbf{k},\mathbf{r})
=  &    2\pi D  \int_{\textrm{FBZ}}
\left|\mathbf{k}-\mathbf{k}' \right|  [1-f_\zs(\mathbf{k})] f_{-\zs}(\mathbf{k}') \times \nn \\& (f_{BE} (\mathbf{k}-\mathbf{k}') +1)    \zd(E_\zs(\mathbf{k})-E_{-\zs}(\mathbf{k}')+ \zo  ) \frac{\dif\mathbf{k}'}{(2\pi)^3}
\end{align*}
where $f_{BE}$ is the Bose-Einstein distribution function, $\zo$ the phonon frequency and $D$ is the deformation potential \cite{Baral_14}. The absorbtion term is obtained by making the substitution  $\zo\rightarrow -\zo $, $f_{BE}+1\rightarrow f_{BE} $.
Finally, the mean electron-phonon collision time  is given by
\begin{align*}
\frac{1}{\tau^{\textrm{em}}_{\zs} (\mathbf{k} ) } =  & 2\pi D  \int_{\textrm{FBZ}}
\left|\mathbf{k}-\mathbf{k}' \right|  [1-f_\zs(\mathbf{k}')] \times \nn  \\& (N (\mathbf{k}-\mathbf{k}') +1)  \zd(E_{-\zs}(\mathbf{k})-E_\zs(\mathbf{k}')+\zo  )\frac{\dif\mathbf{k}'}{(2\pi)^3}\;.
\end{align*}

\section{Acknowledgement}

We thank C. Boeglin, N. Bergeard and O. Bengone for many useful discussions.


\end{document}